\documentclass[a4paper]{a2}
\usepackage{graphicx}
\begin{document}
\def\ea{et~al.}
\def\gi{\; \mathrm{Gyr}}
\newcommand{\rbox}[1]{\raisebox{1.5ex}[0pt]{#1}}

%  \thesaurus{06(06.09.1;
%              06.15.1;02.18.8)}

   \title{The age of the Sun and the relativistic corrections in the EOS}
   \subtitle{}
   \author{A.~Bonanno\inst{1}\and H.~Schlattl\inst{2,3}\and
   L.~Patern\`o\inst{3}}

\offprints{A.~Bonanno, \email{abo@ct.astro.it}}
\institute{INAF - Osservatorio Astrofisico di Catania, Citt\`a Universitaria, 
I-95123 Catania, Italy
	        \and
Astrophysics Research Institute, Liverpool John Moores
University, Twelve Quays House, Egerton Wharf, Birkenhead CH41 1LD,
United Kingdom
                \and
            Dipartimento di Fisica e Astronomia 
dell'Universit\`a, Sezione Astrofisica, Citt\`a Universitaria, I-95123 Catania, Italy}
\date{\today}

\authorrunning{A.~Bonanno et al.}
\titlerunning{The solar age and the relativistic correction}

\abstract{We show that the inclusion of special relativistic
corrections in the revised OPAL and MHD equations of state has a
 significant impact on the helioseismic determination of the solar age.
Models with relativistic corrections included lead to a reduction of about 
$0.05 - 0.08 \gi $ with respect 
to those obtained with the old OPAL or MHD EOS. Our best-fit value is 
$t_\mathrm{seis} = (4.57 \pm 0.11) \gi $ which is in remarkably good agreement
with the meteoritic value for the solar age. 
We argue that the inclusion of relativistic corrections is important for
probing the evolutionary state of a star 
by means of the small frequency separations
$\delta\nu_{{\ell,n}}=\nu_{{\ell,n}}-\nu_{{\ell+2,n-1}}$, 
for spherical harmonic degrees $\ell = 0,1$ and radial order $n \gg
   \ell$.
}
\maketitle

\section{Introduction}
The possibility of using helioseismology to constrain the solar age has been 
discussed by several authors in the past. 
Very recently Dziembowski \ea\ (1999) have shown that the most robust and accurate method is
provided by the small frequency separation analysis (SFSA), 
$\delta\nu_{{\ell,n}}=\nu_{{\ell,n}}-\nu_{{\ell+2,n-1}}$, 
for spherical harmonic degrees $\ell = 0,1$ and radial order $n \gg \ell$ (Tassoul 1980). 

The important property of this quantity is its
strong sensitivity to the sound-speed gradient 
near the solar centre and its weak dependence on
the details of the treatment of the outer layers.  Despite our ignorance of  
a reliable convection model for the solar envelope we are
therefore able to verify how well our models are able to
reproduce the deep radiative regions, in particular the solar core.
Since the properties of the core are mainly determined by
the present central hydrogen abundance, and the latter is influenced
by the solar age, SFSA is a reliable tool to
examine the seismic age of the Sun.

Adopting the OPAL equation of state ( Rogers
\ea\ 1996) a seismic age of  
$(4.66 \pm 0.11) \gi$ has been obtained by
Dziembowski \ea\ (1999), which is consistent with the
meteoritic age $(4.57\pm 0.02) \gi$ of Bahcall \ea\ (1995). 

The aim of this paper is to show that an important ingredient in this
type of analysis is the usage of an accurate equation of state (EOS). 
In particular, by the inclusion of the special relativistic corrections,
like in the updated version of the OPAL EOS, the helioseismic age of
the Sun is reduced to $(4.57\pm 0.11)\ \mathrm{Gyr}$, which is in
remarkable agreement with the meteoritic value. 

Elliott \& Kosovichev (1998) have demonstrated that 
the inclusion of relativistic corrections in 
the EOS leads to a better agreement between the solar models and 
the seismic Sun. By  inverting SOI-MDI/SOHO $p$-mode frequencies 
they found that the solar adiabatic exponent $\Gamma_1$ is much
better reproduced by solar models including the relativistic contribution 
to the Fermi-Dirac statistics. Since the improved EOS 
causes a decrease of $0.2\%$ in the adiabatic index $\Gamma_1$
in the solar centre, the sound speed ($\propto \sqrt{\Gamma_1}$) is
reduced by about 0.1\%. Therefore, the influence of the
relativistic corrections should also be visible in the 
small frequency separations $\delta\nu_{{\ell,n}}$. Indeed,  
Bonanno \ea\ (2001) have found that including this effect in the value
of $\Gamma_1$ improves the
agreement in $\delta\nu_{{\ell,n}}$ between solar models and observations,
thereby confirming the results of Elliott \& Kosovichev (1998).

%%%%%%%%%%%%%%%%%%%%%%% TABLE %%%%%%%%%%%%%%%%%%%%%%%%%%%%%
\begin{table*}
\label{tab1}
\caption{Characteristic quantities of selected solar models. The
indices $0$, $\mathrm{ph}$, $\mathrm{cz}$, and $c$ denote initial,
photospheric, bottom of 
convective envelope, and centre, respectively. MHD-R is the
abbreviation for the MHD EOS  
containing the relativistic corrections in $\Gamma_1$.}
\begin{flushleft}
\begin{tabular}{*{13}{c}}
\hline
\hline
Model &  $\mathrm{\frac{age}{Gyr}}$ &  EOS & $Y_0$& $Z_0$& $Y _\mathrm{ph}$& $Z _\mathrm{ph}$& 
$\frac{r_\mathrm{cz}}{R_\mathrm{ph}}$ & $X_c$&  $Y_c$ & $\frac{\rho_c}{\mathrm{gcm^{-3}}}$ & $\frac{T_c}{10^6\, \mathrm{K}}$ & 
$\frac{S_{pp}(0)}{\mathrm{10^{-25}\;MeV\, b}}$ \\
 \hline  
 1& 4.58&  OPAL 01& 0.2755& 0.01995& 0.2453& 0.01805&  0.7132& 0.3353& 0.6432& 152.87&   15.73 & 3.89\\
 2& 4.58&  OPAL 96& 0.2749&  0.01995& 0.2449& 0.01806&  0.7132& 0.3289& 0.6428& 152.70&   15.72 
& 3.89\\
 3 & 4.60&  OPAL 01 & 0.2752&  0.01995& 0.2451& 0.01805&  0.7125& 0.3342& 0.6443 & 153.16&  15.73 
& 3.89\\
 4& 4.60&  MHD-R& 0.2757&  0.01997& 0.2452& 0.01805&  0.7141& 0.3341& 0.6444 & 153.22&   15.74 
& 3.89\\
 5& 5.00&  OPAL 01& 0.2714&  0.02013& 0.2405& 0.01816&  0.7082& 0.3133& 0.6650& 159.82&   15.84  
& 3.89\\
 6& 4.58&  OPAL 01& 0.2758&  0.01989& 0.2460& 0.01803&  0.7118& 0.3362& 0.6423& 151.35&   15.66  
& 4.00\\
\hline
\end{tabular}
\end{flushleft}
\end{table*}
%%%%%%%%%%%%%%%%%%%%% END TABLE%%%%%%%%%%%%%%%%%%%%%%%%%%%%

In addition to the age, the central hydrogen abundance is also
crucially dependent on the precise value of $S_{pp}(0)$, the zero-energy
astrophysical S-factor for the proton-proton fusion cross section. 
Schlattl \ea\ (1999) and Antia \& Chitre (1999) have shown, using the old
version of the OPAL EOS, that an increase of $S_{pp}(0)$ by about 4\%
with 
respect to Adelberger \ea's~(1998) value yields a
better agreement with the observed frequencies for an age of $4.57$~Gyr.
For this reason we consider in our analysis also different values of
$S_{pp}(0)$.

Including the updated OPAL EOS the best agreement between
meteoritic and seismic age could be achieved with Adelberger
\ea's~(1998) $S_{pp}(0)=4.00 \times 10^{-25}~\mathrm{MeV \,b}$. 
Hence, by taking into account the relativistic corrections in the EOS
there is no need for an artificial increase of $S_{pp}(0)$, as suggested by
previous works, in order to obtain a better agreement between seismic
and meteoritic age.

The code and physics used to compute the various solar models are
described briefly in the next section, followed by the consequences
for the seismic age obtained by means of the SFSA
(Section 3). In the final part the results are discussed.

\section{The new solar models}
We computed a large number of solar models using the GARching
SOlar Model  (GARSOM) code which has been described in its latest
version in Schlattl (2001). 
Our standard model has been compared with other contemporary solar models 
by Turck-Chi\`eze \ea\ (1998), who found a good agreement between
various programs.

The solar photospheric radius and luminosity have been assumed to be
$695.51~\mathrm{Mm}$ (Brown \& Christensen-Dalsgaard 1998) and
$3.8646\times  10^{33} \;\mathrm{erg/s}$, respectively. The surface
metal ratio
has been taken from Grevesse \& Noels (1993), thus $Z/X=0.0245$.
The mixing length parameter (B{\"o}hm-Vitense 1958), initial helium and metal
content have been adjusted in all models to reproduce these values
with an accuracy better than $10^{-4}$.  

In the actual calculations the latest OPAL-opacities (Iglesias \&
Rogers 1996) completed in the low-temperature regime by tables of
Alexander \& Fergusson (1994) have been implemented. 
The outer boundary condition was determined assuming an Eddington grey
atmosphere.
Microscopic diffusion of hydrogen, helium and all major metals is
taken into account. For the EOS we used either the OPAL- (Rogers
\ea\ 1996) or the MHD-tables (Hummer \& Mihalas 1988, Mihalas \ea\ 1988, 
D{\"a}ppen \ea\ 1988). The original OPAL EOS (OPAL96) has been updated by treating
electrons relativistically and by improving the activity expansion
method for repulsive interactions (Rogers 2001), denoted OPAL01 in the
following.

In the case of MHD EOS the relativistic corrections are not directly 
included in the tables. We have therefore corrected the adiabatic index $\Gamma_1$ 
employing the expression of Elliott \& Kosovichev (1998),
\begin{equation}\label{rec}
\frac{\delta\Gamma_1}{\Gamma_1}\equiv {\Gamma_{1, \mathrm{rel}}-\Gamma_{1} 
\over \Gamma_1}\simeq -
\frac{2+2X}{3+5X}\;\frac{k T}{m_e c^2},
\end{equation}
where $T$ is the temperature, $m_e$ the electron mass, 
$c$ the light speed in vacuum, $k$ the Boltzmann constant, 
and $X$ the hydrogen mass fraction. 
As expected, the correction to 
$\Gamma_1$ is negative,  
since its value is 5/3 for the non-relativistic and 4/3
for the extremely relativistic case.

The nuclear reaction rates are taken either from Bahcall \ea\ (1995) or
from  Adelberger \ea\ (1998) with  
$S_{pp}(0)$ being $3.89 \times 10^{-25}~\mathrm{MeV \,b}$
in the first and $4.00 \times 10^{-25}~\mathrm{MeV \,b}$
in the latter case. Other differences in the reaction rates are not
very significant in determining the evolutionary stage of the solar core.

\section{Results for the solar age}
We have computed solar models following the evolution from the zero-age main sequence
with ages ranging from $4.40$ to $5.00$~Gyr in steps of $0.1$~Gyr.
Some basic quantities of a selection of models are summarized in
Table \ref{tab1}. 

%%%%%%%%%%%%%%%%%%%%%%%%%%%%%%%%%%%%%%%%%%%%%%% FIG 1 %%%%%%%%%%%%%%%%%%%%%%%%%
\begin{figure}[t]
\resizebox{\hsize}{!}{\includegraphics{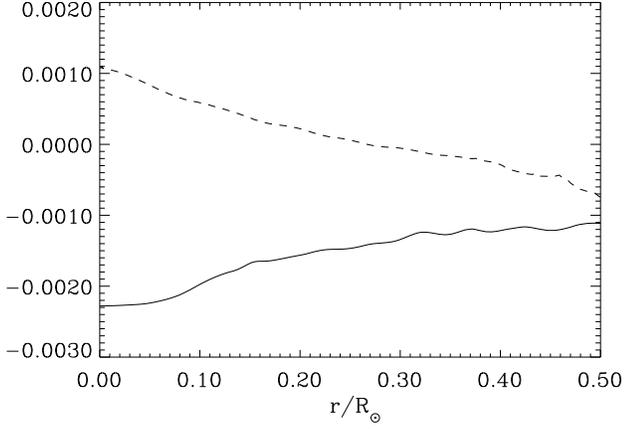}}
\caption{The differences of $\Gamma_1$ (solid line) and $\rho$
(dashed) between two models which either neglect (model 1) or contain the
relativistic corrections (model 2) in the sense $\mathrm{(model\,2 - model\,1)/(model\,1)}$.\label{profile}}
\end{figure}
%%%%%%%%%%%%%%%%%%%%%%%%%%%%%%%%%%%%%%%%%%%%%%%%%%%%%%%%%%%%%%%%%%%%%%%%%%%%%%

For the higher ages the initial helium content has to be reduced to obtain
the correct solar luminosity (compare models 1 and 5). Nevertheless, a
larger lifetime leads to a steeper He profile toward the centre
causing a larger central He abundance. 
The consequent increase of the opacity near the core demands an
higher central temperature to produce the  
same amount of energy. This effect is further
enhanced by 
diffusion which is operating longer for greater ages and is further
increasing the central He content.
Since the relativistic correction to $\Gamma_1$ increases with temperature
(Eq.\ \ref{rec}), the inclusion of relativistic effects has a larger
influence on older models. The
relative differences in the profiles of $\Gamma_1$ and the density are shown
in Fig.~\ref{profile}.  

Models with greater $S_{pp}(0)$, but the same age, have a smaller $T_c$
(see models 1 and 6 in Table~\ref{tab1}), as
the hydrogen burning in the core is more efficient.

In order to determine the seismic age, we calculated for all the
solar models the small frequency separations
$\delta\nu_{{\ell,n}}$ for $\ell = 0,1$ and ${n}\gg\ell$. 
These values have been compared with latest GOLF/SOHO data for $\ell =
0,1,2,3$, which have been obtained from long time series, and where the 
asymmetric line profile has been taken into account during the data reduction 
(Thiery \ea\ 2000). Only the frequencies of the mean multiplet ($m$=0)
are used, as for them the influence of rotation is smallest. 

\begin{figure}[t]
\resizebox{\hsize}{!}{\includegraphics{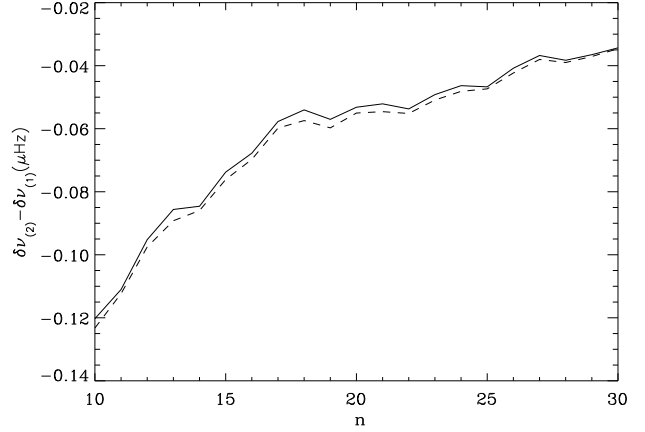}}
\caption{The differences of the quantity $\delta\nu_{n,\ell}$ 
between two models which either neglect (model 1) or contain the
relativistic correction (model 2) in the sense $\mathrm{(model\,2 - model\,1)}$
for an age of $4.20 \gi$ (solid line) and $4.70 \gi$ (dashed line).\label{diff}}
\end{figure}

For the analysis, the $\chi^2$ method has been used, as in
Dziembowski \ea\ (1999) or Schlattl \ea\ (1999);
\begin{equation}
\chi^2_\ell=\frac{1}{ M-m}\sum_{ {n=m}}^{ M}\frac{(\delta\nu_{{\ell,{n}},
\odot}-\delta\nu_{{\ell,{ n,\mathrm{model}}}})^2}{\sigma^2_{{\ell,{n}}}+\sigma^2_{{\ell+2,{n-1}}}} 
\label{eq2}
\end{equation}
with $M=31$ for $\ell =0$
and $M=27$ for $\ell=1$, and $m$ being $10$ in both cases.
It is interesting to notice that including the relativistic
corrections leads to a reduction of $\delta\nu_{{\ell,n}}$ of about
$0.1~\mathrm{\mu Hz}$ for low frequencies (Fig.\ref{diff}).

%%%%%%%%%%%%%%%%%%%%%%%%%%%% FIG 2 %%%%%%%%%%%%%%%%%%%%%%%%%
%%%%%%%%%%%%%%%%%%%%%%%%%%%%%%%%%%%%%%%%%%%%%%%%%%%%%%%%%%%%
\begin{table}[t]
\caption{The best-fit age and the corresponding minimum of $\chi^2$ for 
the grid with different equations of state and different values of $S_{pp}(0)$
in units of $10^{-25}~\mathrm{MeV\,b}$.\label{min}}
\tabcolsep.5em
\begin{tabular}{c|c|cc|cc}
\hline
\hline
& & \multicolumn{2}{c|}{$\ell=0$} & \multicolumn{2}{c}{$\ell=1$}  \\
\rbox{EOS} & \rbox{$S_{pp}(0)$} &$t_\mathrm{seis}$ & 
$\chi_0^2$ & $t_\mathrm{seis}$ & $\chi_1^2$   \\
 \hline
OPAL96 & 3.89 & $4.664\pm 0.088$ & 1.05 & $4.672\pm 0.088$ & 1.66  \\
OPAL01 & 3.89 & $4.584\pm 0.088$ & 1.45 & $4.624\pm 0.072$ & 1.66  \\
MHD & 3.89 & $4.664\pm 0.080$ & 1.00 & $4.680\pm 0.095$ & 1.65  \\
MHD-R & 3.89 & $4.608\pm 0.040$ & 1.07 & $4.640\pm 0.088$ & 1.25  \\
OPAL01 & 4.00 & $4.552\pm 0.080$ & 1.34 & $4.584\pm 0.080$ & 1.47  \\
\hline
\end{tabular}
\end{table}

\begin{figure}[t]
\resizebox{\hsize}{!}{\includegraphics{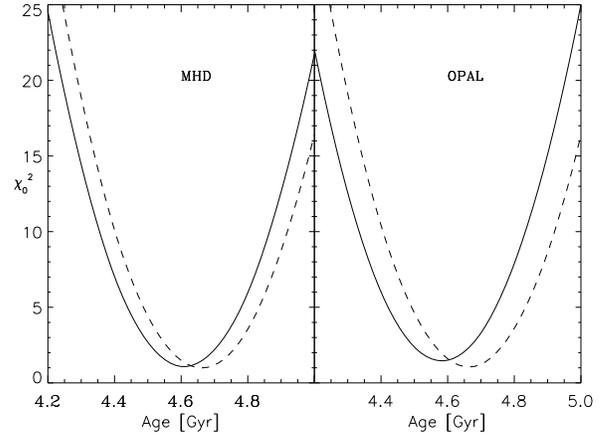}}
\caption{$\chi^2_0$ for models with different age, neglecting (dashed
line) or including the relativistic  correction (solid line). The MHD-EOS
has been used for the models in the left panel, the OPAL-EOS in the ones
of the right panel.\label{chi0}}
\end{figure}

\begin{figure}[t]
\resizebox{\hsize}{!}{\includegraphics{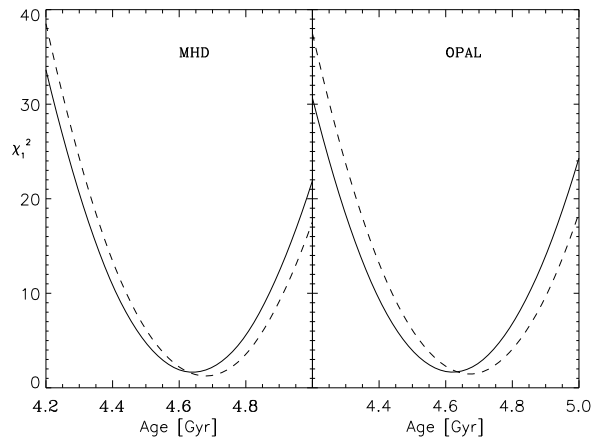}}
\caption{Same as Fig.~\ref{chi0}, but for $\chi^2_1$.\label{chi1}}
\end{figure}

The results for the $\chi^2$-values in models with different ages
are shown in Figs.~\ref{chi0} and \ref{chi1}. The best-fit age 
given by the minimal $\chi^2$-value ($\chi^2_\mathrm{min})$ and the
error determined by the condition $\chi^2 - \chi^2_\mathrm{min} \le 1$
are summarized in Table~\ref{min}.

Regardless of whether MHD or OPAL EOS is used, the best-fit age is reduced
by about 0.05--0.08$\gi$ when the relativistic corrections are included.
The minimal value of $\chi_0^2$ is not significantly
different for all the cases, although the models with OPAL96
EOS have a slightly smaller $\chi^2_{0,\mathrm{min}}$   
than those obtained with OPAL01 EOS. 

It is worth noticing that with $S_{pp}(0)=4.00 \times 10^{-25}~\mathrm{MeV\,b}$
the minimum $\chi^2$-value slightly improves for both $\ell =0$ and
$\ell =1$ (Table~\ref{min}).
Using OPAL01 EOS, which includes the relativistic corrections 
in a consistent way, we obtain in this case as the best-fit age
$t_\mathrm{seis}= (4.57 \pm 0.11)$~Gyr, where we have taken 
the mean of the best-fit value for $\ell=0$ and $\ell=1$. This 
provides our most reliable value for the seismic solar age.
%%%%%%%%%%%%%%%%%%%%%%%%%%%%%%%%%%%%%%%%%%%%%%%%%%%%%%%%%%%%%%%%%%%%%%%%%%%%%%%%%%%%%
\section{Conclusions}
By using updated versions of the OPAL and MHD EOS the seismic age of
the Sun has been redetermined using SFSA with the 
latest GOLF/SOHO data. The important new ingredient in both equations
of state is the
inclusion of the special relativistic corrections. In both cases
almost the same age has been obtained. 

A crucial quantity in the determination of the seismic age is the
proton-proton fusion rate. With the older
versions of the equations of state, a rate about 4\% higher as
the value of Adelberger \ea\ (1998) appears to be favoured, in 
order to obtain a better agreement between seismic and meteoritic
ages. However, with the updated versions of the OPAL and MHD EOS the
seismic age obtained with 
Adelberger \ea's (1998) 
value for $S_{pp}(0)$ is $(4.57 \pm 0.11)$~Gyr, which is 
in excellent agreement with the meteoritic age of $4.57$~Gyr (Bahcall
\ea\ 1995). 

Therefore, the presently favoured value for $S_{pp}(0)$ is $4.00\times
10^{-25}~\mathrm{MeV \, b}$. However, since the uncertainties, in particular,
in the opacities are supposed to be of the order of a few percent,
$S_{pp}(0)$ can only be determined with a similar accuracy by comparing
seismic and meteoritic ages. 

A further source of uncertainty is the centrifugal and magnetic
distortion, but these effects can be neglected for the Sun, as discussed by 
Dziembowski \ea  ~(1999). 

%The centrifugal distorsion can be neglected since
%the relativistic corrections are effective only below $ \approx 0.02$
%solar radii (see Fig. \ref{profile}). In this region the
%ratio between the centrifugal force and the gravitational force is
%of the order of $10^{-6}$ and the shift in the mean 
%multiplet frequencies is about $10~\mathrm{nHz}$, 
%if the second order effects arising from rotational distorsion 
%are taken into account (Antia \ea\ 2001).
%The most important contributions in the sums of Eq.\ref{eq2}
%are from modes up to about $n\approx 20$, 
%and for these terms the relativistic corrections are almost one order
%of magnitude bigger than the rotational distorsion (see Fig. \ref{diff}).
%The magnetic effect is negligible too, because the 
%GOLF data were obtained when the solar activity was near its minimum.

We expect to have asteroseismic data on solar-type 
stars with a precision of about $0.1~\mathrm{\mu Hz}$ from future space missions or 
high-precision ground-based multi-site spectrographic observations. 
We thus think that this effect must be included in the 
standard modelling of solar-like stars when discussing the 
evolutionary changes in the stellar core. 

\acknowledgement{
We thank H.M.Antia for many helpful discussions.
The work of H.S.\ is supported by a Marie Curie
Fellowship of the European Community programme ``Human Potential''
under contract number HPMF-CT-2000-00951.}


\begin{thebibliography}{}
\bibitem[1998]{Adelberger} Adelberger, E.~G., Austin, S.~M., Bahcall,
J.~N., {et~al.} 1998, Rev.\ Mod.\ Phys., 70, 1265
\bibitem[1994]{Alexander} Alexander, D.~R., \& Fergusson, J.~W. 1994, \apj, 437, 879
\bibitem[1999]{antia} Antia, H.~M. \& Chitre, S.~M. 1999, \aap, 347, 1000
\bibitem[2001]{antia2} Antia, H.~M., Chitre, S.~M., \& Thompson, M.~J.,
2001, \aap, 360, 335
\bibitem[1995]{Bah}Bahcall, J.~N., Pinsonneault, M.~H., \& Wasserburg,
G.~J. 1995, Rev.\ Mod.\ Phys., {67}, 781
\bibitem[1958]{MLT}B{\"o}hm-Vitense, E. 1958, Z.~Astrophys., 46, 108
\bibitem[2001]{bonanno1} Bonanno, A., Murabito, A., \& Patern{\`o}, L. 2001, \aap, 375, 1062
%\bibitem[2001]{Bonanno} Bonanno A. \& Patern\`o L. 2001, Mem.~Soc.~Astron.~Ital., 73, 496. 	
\bibitem[1998]{Brown} Brown, T.~M. \& Christensen-Dalsgaard, J. 1998, \apj, 500, L195
\bibitem[1988]{Dappen} D{\"a}ppen, W., Mihalas, D., Hummer, D.~G., \& Mihalas, B.~W. 1988, \apj, 332, 261
\bibitem[1998]{dziem} Dziembowski, W.~A., Fiorentini, G., Ricci, B.,
\& Sienkiewicz, R. 1999, \aap, 343, 990
\bibitem[1998]{Elliot} Elliott, J.~R. \& Kosovichev, A.~G. 1998, \apj,
500, L199  
\bibitem[{Grevesse \& Noels(1993)}]{GN93}Grevesse, N. \& Noels,
A. 1993, Phys.~Scripta, T47, 133
\bibitem[1988]{Hummer} Hummer, D.~G. \& Mihalas, D. 1988, \apj, 331, 794
\bibitem[1996]{Iglesias}Iglesias, C.~A. \& Rogers, F.~J. 1996, \apj, 464, 943
\bibitem[1988]{Mihalas} Mihalas, D., D{\"a}ppen, W., \& Hummer, D.~G. 1988, \apj, 331, 815
\bibitem[{Rogers(2001)}]{Rog01}Rogers, F.~J. 2001, Contrib.~Plasma
Phys., 41, 179
\bibitem[1996]{Rog96}Rogers, F.~J., Swenson, F.~J., \& Iglesias, C.~A. 1996, \apj, 456, 902
%\bibitem[1997]{Schlattl1} Schlattl H., Weiss A., \& Ludwig H.-G. 1997, \aap, 322, 646. 
\bibitem[1999]{Schlattl2} Schlattl, H., Bonanno, A., \& Patern\`o,
L. 1999, \prd, 60, 113002
\bibitem[2001]{Schlattl3} Schlattl, H. 2001, \prd, 64, 013009
\bibitem[1980]{Tassoul} Tassoul, M. 1980, \apjs, 43, 469
\bibitem[2000]{Thiery} Thiery, S., Boumier, P., Gabriel, A.~H., {et al.}
2000, \aap, 355, 743
\bibitem[1998]{gong98}
  Turck-Chi{\` e}ze, S., {Basu}, S., {Berthomieu}, G., {et~al.} 1998, in
  Structure and Dynamics of the Interior of the Sun and Sun-like Stars, ed.
  S.~Korzennik \& A.~Wilson, ESA SP-418 (ESA Publication Division, Noordwijk,
  The Netherlands), 555
\end{thebibliography}
\end{document}